\documentclass[aps,prl,article,twocolumn,preprintnumbers,amsmath,amssymb,superscriptaddress]{revtex4-2}
\usepackage[utf8]{inputenc}
\usepackage{float}
\usepackage{commath}
\usepackage{graphicx}
\usepackage{verbatim}
\usepackage{color}
\usepackage{physics}
\usepackage[dvipsnames,svgnames,x11names,hyperref]{xcolor}
\usepackage[normalem]{ulem}
\usepackage{algorithm}
\usepackage{algpseudocode}
\usepackage{times}
\usepackage[compat=1.1.0]{tikz-feynman}\tikzfeynmanset{warn luatex=false}
\usepackage[toc,page]{appendix}
\usepackage{siunitx}
\usepackage{dcolumn}   
\usepackage{bm}        
\usepackage{amssymb}   
\usepackage{amsmath}
\usepackage{epigraph}
\usepackage{braket}
\usepackage{gensymb}
\usepackage{ tipa }
\usepackage{bbold}
\usepackage{esint}
\usepackage{mathdots}
\usepackage{float}
\usepackage{comment}
\usepackage[colorlinks=true, linkcolor=blue, urlcolor=blue, citecolor=blue]{hyperref}
\DeclareSIUnit\gauss{G}

\definecolor{linkcolor}{RGB}{6,69,173} 
\definecolor{diffcolor}{RGB}{175,31,36} 

\newcommand{\up}{\uparrow}
\newcommand{\dn}{\downarrow}

\allowdisplaybreaks

\begin{abstract}
We investigate the Nagaoka-Thouless ferromagnetic instability in the strongly interacting $t$-$t'$ Hubbard model by continuously breaking particle-hole symmetry on a tunable square-triangular lattice geometry. We use an analytic approach to show that the fully spin-polarized state becomes unstable to a metastable spin-polaron when the kinetic frustration $t'/t$ exceeds a critical, dimension-dependent value. Large-scale density matrix renormalization group simulations reveal a quantum phase transition from the Nagaoka ferromagnet to a spiral spin-density wave, which evolves continuously into the Haerter-Shastry antiferromagnet in the large-frustration limit. Remarkably, this transition remains robust at low but finite hole density, making it accessible in cold-atom and moiré Hubbard platforms under strong interactions. A variational analysis further captures the instability mechanism at finite hole density via frustration-induced magnon band deformation.
\end{abstract}

\makeatletter
\newcommand{\printauthorsagain}{
  \begingroup
  \let\@author\@empty
  \let\@affiliation\@empty
  \let\@email\@empty
  \@author
  \endgroup
}
\makeatother

\begin{document} 
\title{Instability of Nagaoka State and Quantum Phase Transition via Kinetic Frustration Control}

\author{Prakash Sharma}
\email{sharmaprakash078@gmail.com}
\affiliation{Department of Chemistry, Emory University, Atlanta, Georgia 30322, USA}
\affiliation{Department of Physics and Astronomy, California State University, Northridge, California 91330, USA}

\author{Yang Peng}
\affiliation{Department of Physics and Astronomy, California State University, Northridge, California 91330, USA}
\affiliation{Institute of Quantum Information and Matter and Department of Physics, California Institute of Technology, Pasadena, CA 91125, USA}

\author{Donna N. Sheng}
\affiliation{Department of Physics and Astronomy, California State University, Northridge, California 91330, USA}

\author{Hitesh J.  Changlani}
\affiliation{Department of Physics, Florida State University, Tallahassee, Florida 32306, USA}
\affiliation{National High Magnetic Field Laboratory, Tallahassee, Florida 32310, USA}

\author{Yao Wang}
\affiliation{Department of Chemistry, Emory University, Atlanta, Georgia 30322, USA}

\maketitle

\textit{Introduction---}Itinerant ferromagnetism (FM), proposed by Nagaoka and Thouless\,\cite{nagaoka1966ferromagnetism,thouless1965exchange}, arises when the motion of an itinerant electron kinetically stabilizes ferromagnetic order in an almost half-filled correlated system, which is remarkably distinct from the superexchange mechanism or the Stoner mechanism based on a mean-field instability of an interacting Fermi sea~\cite{stoner1938collective}.
Beyond its unconventional origin, this mechanism enables direct electrical control of magnetism, as doping modulates global magnetic order by altering carrier kinetics~\cite{sharpe2019emergent,kittilstved2006direct}. It also serves as an efficient measurement for strong electronic correlations, offering a sensitive probe of the Hubbard interaction strength in moiré materials\,\cite{mak2022semiconductor,wu2018hubbard} where high-resolution single-particle spectra are inaccessible~\cite{nuckolls2024microscopic,xie2022strong}. Furthermore, the physics of kinetic FM bridges to the double exchange mechanism observed in strongly correlated oxides\,\cite{zener1951interaction,tokura2000orbital,dagotto2001colossal}, providing a unified perspective on doping-induced magnetism. While originally limited to extreme conditions in the Hubbard model, generalized forms of kinetic FM have recently been successfully realized at finite doping and experimentally accessible energy scales in moiré heterostructures\,\cite{tang2020simulation,ciorciaro2023kinetic} and optical lattices\,\cite{prichard2024directly,lebrat2024observation,koepsell2019imaging}.

\begin{figure}
\includegraphics[width=\linewidth]{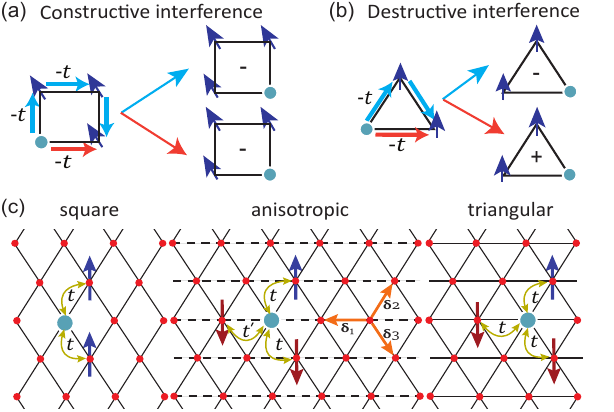}
    \caption{
    Path interference of a hole in (a) a square and (b) a triangular plaquette within a FM background. On the square lattice, two hopping paths (light blue and red arrows) yield the same final configuration with identical phase and interfere constructively.
    On the triangular lattice with $t>0$, two paths yield final states that differ by a fermionic sign and interfere destructively. (c) Lattice geometry interpolating between square and triangular limits. The unfrustrated square lattice has nearest-neighbor hopping $t$, while introducing diagonal hopping $t'$ creates kinetic frustration. The fully frustrated triangular lattice is recovered when all bonds have equal hopping amplitude $t$. The nearest neighbor vectors $\boldsymbol{\delta_1}$, $\boldsymbol{\delta_2}$ and $\boldsymbol{\delta_3}$ of triangular lattice are shown. 
}
\label{fig:fig1}
\vspace{-0.1in}
\end{figure}

In contrast to the square lattice geometry where Nagaoka ferromagnetism (NFM) was originally predicted, recent studies have indicated the triangular lattice as a particularly compelling setting for exploring itinerant magnetism due to its inherent geometric frustration, rich phase diagram with particle-hole asymmetry~\cite{lee2023triangular,morera2023high,samajdar2024nagaoka,xu2023frustration,li2014competing,pereira2025kinetic}, and the realizability in moiré superlattices~\cite{tang2020simulation,zhang2023pseudogap}. In bipartite (e.g., square) lattices, a hole's kinetic energy is minimized by constructive interference along hopping pathways [Fig.\ref{fig:fig1}(a)], favoring uniform spin alignment and stabilizing a conventional NFM phase\,\cite{nagaoka1966ferromagnetism,thouless1965exchange}. In contrast, triangular lattices host closed loops with an odd number of bonds, inherently introducing destructive interference for certain hopping pathways [Fig.\ref{fig:fig1}(b)], suppressing fully spin-polarized Nagaoka order at infinitesimal doping~\cite{haerter2005kinetic,sposetti2014classical} and stabilizing intermediate states such as spin polarons\,\cite{martinez1991spin,davydova2023itinerant,zhang2018pairing,newby2025finite}. Recent experiments with optical lattices and moiré heterostructures have explicitly observed a spin-polaron to local Nagaoka-like ferromagnetic transition driven respectively by hole and particle (doublon) dopants\,\cite{lebrat2024observation,prichard2024directly,koepsell2019imaging,ciorciaro2023kinetic,tao2024observation}, revealing fundamentally different doping dependence and transition characteristics from those observed in bipartite systems. 

Despite the difference, the bipartite square lattice and geometrically frustrated triangular lattice can be gradually connected to each other by tuning one of the diagonal hopping $t'$ in a square lattice [see Fig.\ref{fig:fig1}(c)]. 
Since the kinetic frustration~\cite{barford1991spinless,merino2006ferromagnetism} of a hole dopant in the fully frustrated triangular limit is alleviated when the background spins adopt antiferromagnetic correlations arranged in a $120^\circ$ order\,\cite{haerter2005kinetic}, the fundamental distinction between these two systems and their magnetic origin indicate a geometric phase transition through this lattice interpolation and potentially richer magnetic phases distinct from either ends.

To this end, we systematically investigate the instability of the Nagaoka-type ferromagnet and its evolution toward the low-spin antiferromagnet by continuously tuning lattice geometry from square to triangular. A similar problem has been previously explored using numerical simulations for a single-hole dopant system~\cite{lisandrini2017evolution}. In this study,  we combine an analytical solution with large-scale DMRG to elucidate the instability mechanism and the underlying ground state phase transition in both the single-hole-dopant and finite-hole-density cases.
Our single-hole analysis also reveals that Tasaki's criteria for generalized NFM\,\cite{tasaki1989extension} are sufficient but not necessary. 
Most strikingly, our DMRG results beyond the idealized single-hole limit uncover that critical quantum phenomena linking NFM and the Haerter-Shastry ($120^\circ$ AFM) phase~\cite{haerter2005kinetic}, driven by competing magnetic correlations, persist at finite hole doping, making them directly relevant to cold-atom and moiré Hubbard experiments. We explain the physical origin of the instability of NFM at finite hole density using a simple yet insightful variational analysis, revealing that the instability emerges directly from the frustration-induced modification of dispersive magnon bands. 

\textit{Geometric control of kinetic frustration---}To study the magnetism of purely kinetic origin, the $U=\infty$ Hubbard model is an ideal platform, as magnetic interactions ($J\sim t^2/U$) exactly vanish in this limit, and the model reduces to a constrained nearest-neighbor hopping,
\vspace{-0.07in}
\begin{equation}
    \mathcal{H} = -\sum_{\langle i,j \rangle,\sigma} t_{ij} c_{i\sigma}^\dagger c_{j\sigma} \left(1-n_{j-\sigma}\right)
    \left(1-n_{i-\sigma} \right) + \text{h.c.},
    \label{eq:infU_H}
    \vspace{-0.08in}
\end{equation}
where $c_{i\sigma}$ $(c_{i\sigma}^\dagger)$ annihilates (creates) electron at site $i$ with spin $\sigma=\{\uparrow, \downarrow \}$ and the projection operator $(1-n_{i-\sigma})$ prevents from having configurations with two electrons on the same site.
Consequently, electrons can only exchange positions with holes on adjacent sites.

We first consider a single-hole dopant in an otherwise half-filled 2D triangular lattice with hopping strength $t_{ij}= t^\prime$ along horizontal bonds and $t$ otherwise [see Fig.~\ref{fig:fig1}(c)]. We extend the analysis to finite hole density in a later section.
In the infinite-$U$ limit, the ratio $t^\prime/t$ is the only tuning parameter that controls the degree of kinetic frustration, which is equivalent to a change in the lattice geometry: $t^\prime=0$ corresponds to a bipartite (square) lattice with particle-hole symmetry (PHS), where a single-hole-doped state exhibits NFM regardless of the sign of $t$.
In contrast, $t^\prime/t=1$, with $t>0$, corresponds to the fully frustrated triangular lattice. Such a lattice manipulation via tuning the hopping term has already been realized in cold-atom optical simulators by introducing an imbalance between two orthogonal retro-reflected laser beams\,\cite{xu2023frustration}.

As parameter $t'$, which breaks PHS continuously when tuned between $0$ and $t$, introduces kinetic frustration for the hole, it raises the question of how the NFM ground state evolves in response to this frustration. A hole generally acquires different hopping phases depending on the sign and strength of $t'$ (assuming $t=1$ fixed). For example, when $t'<0$, NFM continues to persist due to the product of the hopping phase around a triangular loop being positive\,\cite{merino2006ferromagnetism}. This extends NFM to broader graph structures beyond bipartite geometries, as generalized by Tasaki in Ref.\,\cite{tasaki1989extension}. The more intriguing regime occurs when $t'>0$, where a FM state competes with AFM correlations as $t'$ increases. In this scenario, a hole may favor one or more spin flips along its path to reduce the kinetic frustration by accumulating phases that can effectively modify its hopping\,\cite{sposetti2014classical}.

For $t^\prime>0$, it is not \textit{a priori} obvious whether the FM state can survive in the ground state due to competing many-body effects. More formally, each matrix element satisfies $(\mathcal{H})_{ij}\geq 0$ under an appropriate gauge choice (see Supplementary Material (SM) for details). This violates Tasaki's criterion for NFM, which requires strictly non-positive hopping matrix elements to ensure the fully spin-polarized ground state\,\cite{tasaki1989extension}. Instead, in this regime, the Perron-Frobenius theorem\,\cite{pillai2005perron} guarantees a unique ferromagnetic state (apart from its trivial $N$-fold degeneracy) atop the many-body spectrum, but not necessarily as the ground state.

To analyze this, we investigate the instability of NFM against $t'$ by introducing a single spin-flip into the fully polarized background. This simple ansatz provides a necessary condition for ferromagnetic instability, though it does not guarantee the true ground state--which we revisit later through full many-body DMRG calculations.
The exact wavefunction of the single-hole Nagaoka state is represented by \( \ket{\psi}_{1H} = \sum_j \alpha_j c_{j\uparrow} \ket{\text{FM}} \) (for some amplitude $\alpha_j$) and that of the spin-flip (or one-hole-one-magnon, 1H1M) state by
\vspace{-0.08in}
\begin{equation}
    \ket{\psi}_{1H1M} = \sum_{m \neq n} \alpha_{mn} c_{m\uparrow} S_n^- \ket{\text{FM}},
    \label{eq:1h1m}
    \vspace{-0.1in}
\end{equation}
for some scalar coefficients $\alpha_{mn}$,
where $S_n^-=c_{n\dn}^\dagger c_{n\up}$ and $\ket{\text{FM}}=\prod_{i=1}^N c_{i\up}^\dagger \ket{0}$, with $N$ being the number of sites. The condition $m \neq n$ ensures that a hole and a spin-flip do not occupy the same site. The instability is characterized by the hole–magnon binding energy, \(E_b = E_{\text{1H1M}} - E_{\text{1H}} \), where \( E_{\text{1H}} \) and \( E_{\text{1H1M}} \) are the lowest energies of the 1H and 1H1M states, respectively. When \( E_b < 0 \), the FM state becomes unstable, and the spin-flip state emerges as an energetically favorable metastable configuration; otherwise, NFM remains stable and the binding energy must vanish, i.e., \( E_b = 0 \).

\begin{figure}
    \centering
    \includegraphics[width=\linewidth]{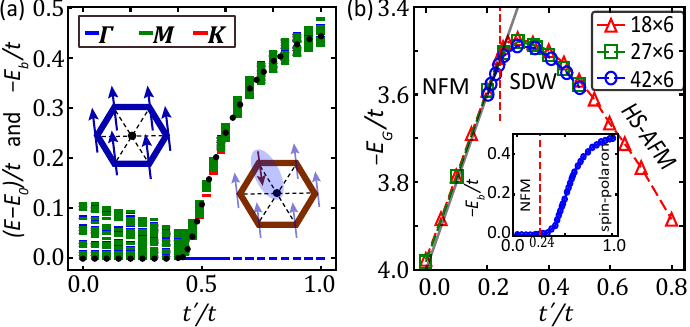}
    \caption{
    (a) Exact low-lying energy spectra (dashed lines) of the hole-magnon system on $80\times80$ torus, shown relative to the ground state energy $E_0$, as a function of $t'/t$ for three center-of-mass momenta: $\mathbf{\Gamma}=(0,0)$, $\mathbf{M}=(0,2\pi/\sqrt{3})$, and $\mathbf{K}=(4\pi/3,0)$. Black dots indicate the hole-magnon binding energy ($-E_b$) on the same plot. Insets depict two competing regimes: a hole favoring Nagaoka FM (left), and a bound spin-polaron state emerging with increasing kinetic frustration (right).  
    (b) Phase diagram and the ground state energy of a hole vs. $t'/t$ for three different width-6 cylindrical clusters. The solid gray line shows the exact FM energy in the thermodynamic limit.  Inset: hole-magnon binding energy on the same width-6 cylinders. Vertical red dashed lines mark the ground state and spin-polaron transitions, both occurring near $t'/t\approx0.24$.
   }
    \label{fig:fig_2}
    \vspace{-0.08in}
\end{figure}

The 1H1M problem reduces to an effective tight-binding Hamiltonian in the relative coordinate (see SM).
Building on the self-consistent method developed for the triangular lattice in Ref.\,\cite{davydova2023itinerant}, we extend it to the $t$-$t'$ model and identify a hole-magnon bound state (spin polaron) that emerges above a dimension-dependent critical value $t_c^\prime$.
In 2D, the bound state at center-of-mass (COM) momentum $\mathbf{P}=\mathbf{0}$ follows from the inversion-odd representation and is given by (see derivations in SM):
\vspace{-0.05in}
\begin{align}
    \left[ 1 + t^\prime f_{33}\right] \left[ 1+ t \left(f_{11}  +  f_{12} \right)\right]
    = 
    2t t^\prime f_{13}^2,
    \label{eq:solution_2D}
    \vspace{-0.12in}
\end{align}
where $f_{ij} = \frac{2}{N}\sum_{\boldsymbol{k\in1BZ}} \frac{\sin(\boldsymbol{k\cdot \delta_i})\sin(\boldsymbol{k\cdot \delta_j})}{E-\epsilon_{\boldsymbol{k}}} $ {\color{black}{is related to Fourier transform of coefficients $\alpha_{mn}$}}, and $\epsilon_\mathbf{k}=\sum_{\boldsymbol{\delta}} t_{\boldsymbol{\delta}} \cos{({\mathbf{k}\cdot \boldsymbol{\delta}})}$ is the dispersion of bare hole (i.e. 1H) state, with $\boldsymbol{\delta}$ representing three neighboring vectors in triangular lattice [see Fig~\ref{fig:fig1}(c)]. 
In the thermodynamic limit $N\rightarrow\infty$, Eq.~\eqref{eq:solution_2D} contains bound state solutions with energy $E$ below the bare hole minima ($\epsilon_\mathbf{k}^{min}$) for  $t^\prime > t_c^\prime \approx 0.42t$. To visualize this, we calculate the many-body eigen energy spectra and hole-magnon binding energy $E_b$ via exact diagonalization on a $80\times80$ torus. For $t^\prime < t_c^\prime$, $E_b=0$, indicating no bound state, while for $t'>t_c'$, negative $E_b$ signals the spin-polaron formation [see Fig~\ref{fig:fig_2}(a) (black dots)], validating our analytic solution. This suggests that despite PHS and Tasaki's condition being broken, NFM remains stable up to a surprisingly large kinetic frustration. 
Above this critical value, the previously unbound hole-magnon pair localizes, with the bound spatial extent shrinking gradually as $t'$ increases towards the triangular limit (see details in SM).  
The bound state is strongest at $t'=1$, with $E_b\approx0.42t$, consistent with Ref.\,\cite{davydova2023itinerant}.

The 1H1M low-lying eigen spectra [dashed lines in Fig.~\ref{fig:fig_2}(a)] further illustrate how kinetic frustration destabilizes NFM, forming a spin-polaron state. 
In the FM regime, $t'<t_c'\approx 0.42t$, magnon spectra appear as a continuum due to excitation gap scaling as $1/N$.  
However, above the critical $t_c'$, the lowest energy excitation is no longer a gapless magnon, but a composite spin-polaron, which introduces a finite gap. 
The lowest energy state shifts from COM $\mathbf{M}$ at small $t'$ to $\mathbf{\Gamma}$ in the spin-polaron regime (also see SM).

\textit{Frustration induced ground state phase transition---}
To identify the true ground state at $U=\infty$, we now focus on the lowest $|\text{S}_z|$ sector. 
Using DMRG, we simulate width-6 cylinders with length along open boundary up to $L_x = 42$, allowing us to access longer-range correlations and reduce finite-size effects.  The simulations are performed using the ITensor open-source library~\cite{ITensor}.
In Fig.~\ref {fig:fig_2}(b), we show ground state energies for three different cylinder lengths.
It reveals a clear transition in the energy trend from linear to quadratic at $t' \approx 0.24t$, indicating a phase transition.
The energy of the fully polarized state in the thermodynamic limit follows $E_G=2t'-4t$, which our data closely tracks for all clusters for $t'<0.24t$, with only minor finite-size effects. We independently verify this phase boundary by examining the spin-polaron transition within the 1H1M sector on the same width-6 cylinders. As shown in the inset of Fig.~\ref{fig:fig_2}(b)], the results are in excellent agreement with the determined ground-state transition point.
Although this transition point, \( t' = 0.24t \), is close to the extrapolated thermodynamic limit found in a previous study~\cite{lisandrini2017evolution}, it is significantly smaller than our estimated 2D limit (\( \sim 0.42t \)) from the single spin-flip picture.
We notice that for any finite-width cylinders, $t_c'$ of the single hole-magnon system decreases as the cylinder width narrows and vanishes in the zigzag ladder limit (i.e., width-2 cylinder), which reflects a dimensional dependence of the Nagaoka state (see details in SM). This situation is similar to the transition between stable and metastable Nagaoka FM with a change in the topology from 2D to 1D, observed in quantum dot array~\cite{ivantsov2020stable}, with a notable difference that the ground state in the zigzag ladder is a spin singlet ($S_{tot}=0$), whereas in the 1D chain it is degenerate with the FM state.

\begin{figure}
\includegraphics[width=\linewidth]{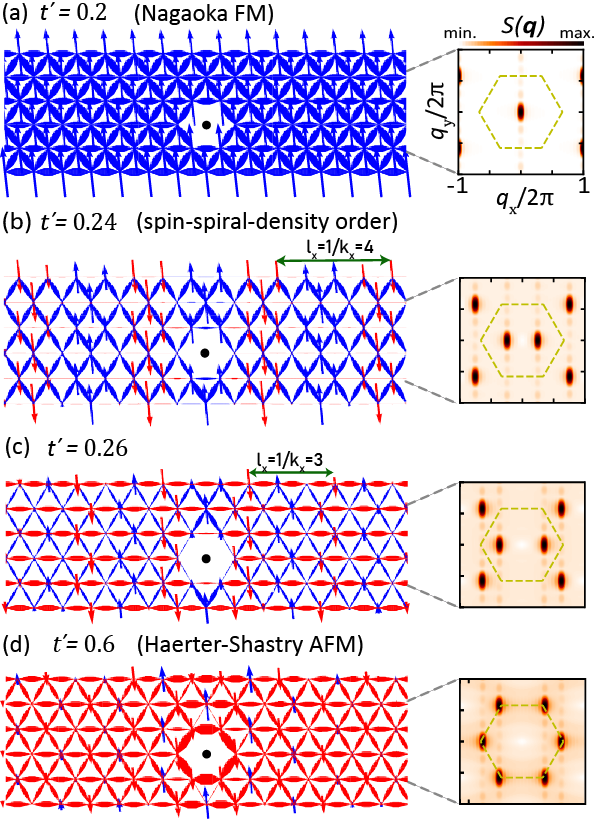}
\caption{Spin correlations profile and structure factor $S(\mathbf{q})$ for: (a) Nagaoka FM at $t'= 0.2t$, (b) spiral-spin--wave at the transition point $t'=0.24t$, (c) spiral order at $t'=0.26t$, and (d) Haerter-Shastry AFM at $t'=0.6t$. Up/down arrows indicate spin correlation $\langle \mathbf{S}_i\cdot \mathbf{S}_o\rangle$ relative to a fixed reference site (black dot); arrow length indicates magnitude, color indicates sign (blue: positive, red: negative). Bond thickness shows quantum snapshots of spin correlations between neighboring sites obtained by freezing the hole at the center (black dot). Bond colors follow the same sign convention.
The dashed hexagon in $S(\mathbf{q})$ represents the Brillouin zone of the triangular lattice. All results are for one hole in $S_z=1/2$ sector, computed via DMRG on an $18\times6$ cylinder with bond dimension up to $D=14000$ and truncation error below $2\times 10^{-6}$.
} 
\label{fig:fig_3}
\vspace{-0.15cm}
\end{figure}
%

To pin down the ground state phase at various $t'$, we calculate the spin correlation $C^{ss}_{ij}=\langle \mathbf{S}_i \cdot \mathbf{S}_j \rangle$ and the static spin structure factor: 
$S(\mathbf{q})=\frac{1}{N}\sum_{i,j} e^{i\mathbf{q}\cdot (\mathbf{r}_i - \mathbf{r}_j)} \langle \mathbf{S}_i \cdot \mathbf{S}_j \rangle$.
In Figs.~\ref{fig:fig_3}(a)-(d), we show the real-space profile of $C^{ss}_{ij}$, represented by up and down arrows, with the reference site fixed at the center of the cluster (black dot). The color (and up/down orientation) of each arrow indicates the sign (blue for positive, red for negative), while the arrow length reflects the magnitude of $C^{ss}_{ij}$. The corresponding $S(\mathbf{q})$ is shown in the rightmost panel of each figure. At $t'/t=0.2$ [Fig.~\ref{fig:fig_3}(a)], below the transition point, spin correlations are saturated to their maximum value $\sim \frac{N-1}{4N}$ (each arrow length is  $\sim0.247$) and the Bragg peak appears at the center of the Brillouin zone (BZ) $\boldsymbol{\Gamma}$  with the peak amplitude close to the expected value $S(\boldsymbol{\Gamma})\approx 26.99$, corresponding to a saturated FM on the $18\times6$ cluster. This NFM survives until the transition point, as evident from the linear energy fit in Fig.~\ref{fig:fig_2}(b). At the transition point, $t'/t=0.24$, the competing FM and AFM correlations lead to an alternating-sign pattern in the spin correlation $C^{ss}_{ij}$, forming multiple ferromagnetic domains [Fig.~\ref{fig:fig_3}(b)]. 
This critical state is a spiral spin wave order with a wavelength of $l_x\approx4$ lattice constants in our finite-width geometry. The associated Bragg peak splits to the ordering vectors $\boldsymbol{q}\approx(\pm \pi/2,0)$, perfectly consistent with the spiral pitch with wavelength being $l_x\approx 4$. As $t'$ is increased above $t_c'$ [Fig.~\ref{fig:fig_3}(c)], the spiral spin wavelength reduces and associated Bragg peaks gradually evolves towards the BZ corners, approaching $\sqrt{3}\times\sqrt{3}$ chiral spin order in the Haerter-Shastry AFM regime at $t'=0.6t$ and above [Fig.~\ref{fig:fig_3}(d)].

Local magnetic bond correlations around a dopant are often useful in cold-atom experiments at finite, but large, $U$ in understanding competing magnetic  orders~\cite{lebrat2024observation,prichard2024directly}. In the large $U$ limit, the superexchange mechanism is suppressed near the dopant, and kinetic effects in the vicinity of the hole become locally accessible. 
To probe this local competition between FM and AFM correlations, we also show the normalized hole-spin-spin correlation $C_{ij}^{hss}=\langle h_0^\dagger \boldsymbol{S}_i\cdot\boldsymbol{S_j}h_0\rangle/\langle{h_0^\dagger h_0\rangle}$ in the ground state, where $h_0 = c_{0\dn} c_{0\dn}^\dagger c_{0\up} c_{0\up}^\dagger$ pins a hole at the central site ``0".
Figs.~\ref{fig:fig_3}(a)–(d) show the nearest-neighbor $C^{hss}_{ij}$, visualized using bond thickness to indicate magnitude and color (blue for FM, red for AFM) to indicate sign, with the hole frozen at the center of the cluster (black dot).
At $t'/t=0.2$, all bond correlations are positive, as expected for NFM. 
However, at the transition point $t'/t=0.24$, spin correlations are FM along each square plaquette, while correlations along horizontal bonds vanish. This critical behavior indicates the onset of the kinetic frustration effect as increasing $t'$ introduces competing hole paths that frustrate the FM alignment and prevent all bonds from being simultaneously satisfied. As $t'$ is increased further, these horizontal bonds reverse sign and become antiferromagnetic [Fig.~\ref{fig:fig_3}(c)].
Enhanced AFM bonds around the dopant in Fig.\ref{fig:fig_3}(d) reflect the hole’s growing tendency to favor local AFM alignments under stronger kinetic frustration.

\textit{Finite hole doping effects---}
After gaining insights from a single-hole dopant, we now investigate finite-hole doping as it is more relevant to experiments. 
{\color{black}{While the NFM state on a square lattice is often debated as a singular-hole artifact, recent analytical and numerical evidence suggests it can persist as a robust thermodynamic phase at finite doping. Rigorous results for the two-hole sector~\cite{tian1991nagaoka} establish a mathematical foundation for finite hole NFM in the thermodynamic limit, while state-of-the-art many-body simulations—including unbiased DMRG and fermionic quantum Monte Carlo—consistently predict a persistent ferromagnetic ground state up to a critical doping of $\delta \approx 0.2$~\cite{carleo2011itinerant,liu2012phases,zhang1991quantum}. The potential for such states to manifest in quantum dots and cold-atom systems gives these findings significant experimental weight, regardless of whether the effects are truly thermodynamic or limited to large, finite-sized clusters.  Despite these advances, the resilience of NFM against the competing effects of kinetic frustration remains largely unexplored.}}

\begin{figure}
    \centering
    \includegraphics[width=\linewidth]{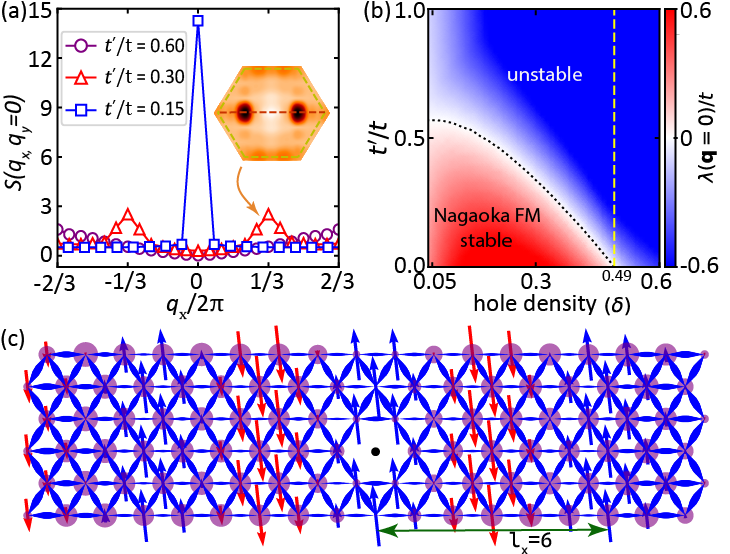}
    \caption{(a) Ground state $S(\mathbf{q})$ along $\mathbf{q_y}=0$ (marked by the horizontal red dashed line in the inset), obtained by considering bulk $72$ sites, for $12$ holes on $18\times 6$ cylinders. Inset: full $S(\mathbf{q})$ within the first BZ. (b) Excitation energy at band bottom $\mathbf{q}=(0,0)$ as a function of hole density and anisotropy $t'/t$, obtained from the SKA variational solution. The dotted black line represents the boundary below which the Nagaoka phase is stable. The yellow dashed line at $\delta=0.49$ indicates the critical density above which NFM is unstable.  
    (c) The real space spin and bond correlation profile for $6$ holes on $18\times6$ cluster near transition at $t'=0.25t$. Plotting conventions follow Fig.~\ref{fig:fig_3}. The size of the purple dots indicates the strength of hole-hole correlations relative to the central site (black dot). 
    }
    \label{fig:multi_hole}
\end{figure}

We use DMRG simulation to characterize the ground state and its evolution under $t'$ hopping at hole densities $\delta\approx0.06$ ($6$ holes) and $\delta\approx0.11$ ($12$ holes) on $18\times6$ cylinders. Finite doping poses significant convergence challenges, requiring bond dimensions up to $D=25000$ with $U(1)$ symmetry to control truncation errors below $1\times10^{-5}$. 
In Fig.~\ref{fig:multi_hole}(a), we show the $S(\mathbf{q})$ with cut along $q_y=0$ for $\delta\approx0.11$ at representative values of $t'$. At $t'=0.15t$, the Bragg peak at $\mathbf{\Gamma}$ (i.e., $q_x=0$) reaches a saturation value $S(\mathbf{\Gamma})\approx14.66$ for the bulk $72$ sites of the cluster, confirming NFM.
At $t'=0.3t$, the Bragg peak splits, similar to the single-dopant case, forming spiral spin density order with ordering vector $q_x=\pm 2\pi/3$. This corresponds to the real-space periodicity $l_x=3$. However, this wavelength is longer compared to the single-hole dopant case, indicating a modified magnetic response at finite doping. A similar effect appears at low doping  $\delta\approx0.06$ [Fig~\ref{fig:multi_hole}(c)], where the periodicity extends to $l_x=6$ near the transition. This longer periodicity arises from the effective repulsion between holes, as seen in the hole-hole correlation (purple dots), where each hole prefers its own FM background, avoiding overlap with others. This is evident from the vanishing hole correlation along the central stripe (blue arrows) where a reference hole is fixed at the center (black dot). 
At both doping $\delta$, we find $120^\circ$ AFM states with peaks at $q_x=\pm 4\pi/3$ at larger $t'$ values, akin to the single-hole dopant case.

Due to the lack of an exact solution for finite hole density, even in the simplest scenario of a single spin-flip, we adopt the variational wavefunction approach introduced by Shastry, Krishnamurthy, and Anderson (SKA)\,\cite{shastry1990instability} to elucidate further the physical origin of instability NFM due to kinetic frustration. Originally, the SKA method was developed to analyze the ``$k_F$ instability" of NFM under finite doping.
While the SKA approach is known to overestimate Nagaoka stability, it effectively captures the key instability mechanism, offering valuable insights where direct numerical simulations are computationally demanding across multiple $\delta$ and $t'$.
The SKA wavefunction describes the leading instability as an overturned (spin-flip) electron from the Fermi surface of the fully polarized state and creating a down-spin electron at the band bottom:
\vspace{-0.08in}
\begin{equation}
    \ket{\psi_v(\mathbf{q})} = \frac{1}{\sqrt{N}} \sum_j e^{i\boldsymbol{q}\cdot\boldsymbol{r}_j} c_{j\downarrow}^\dagger(1-n_{j\uparrow})c_{\mathbf{k}_F\uparrow}\ket{F},
    \label{eq:ska_wavefunction}
    \vspace{-0.1in}
\end{equation}
where $\ket{F}=\prod_{0\le |\mathbf{k}|\le\mathbf{k}_F} c^\dagger_{\mathbf{k}\uparrow}\ket{\text{vac}}$ is the FM state at finite hole density, and $\mathbf{k}_F$ is one of the Fermi surface vectors. 
The projector $1-n_{j\uparrow}$ in Eq.~\eqref{eq:ska_wavefunction} enforces the $U=\infty$ constraint in real space. This allows us to switch off $U$ and leverage the Hamiltonian to non-interacting model, $H=-\sum_{\langle i,j\rangle} t_{ij} c_{i\sigma}^\dagger c_{j\sigma}$. The excitation energy is then given by \vspace{-0.11in}
\begin{equation}
\lambda (\mathbf{q})=\frac{\bra{\psi_v}(H-E_o)\ket{\psi_v}}{\langle \psi_v\ket{\psi_v}},
\vspace{-0.02in}
\end{equation}
where $E_o$ is the energy of the polarized state. 
A straightforward but lengthy calculation gives,
\vspace{-0.1in}
\begin{equation}
    \lambda(\mathbf{q}) \mkern-6mu= \mkern-6mu \mu \mkern-3mu - \mkern-3mu \epsilon_{\mathbf{k}_F} \mkern-3mu + \mkern-2mu \epsilon_\mathbf{q} \delta \mkern-3mu
    + \mkern-3mu \frac{1}{\delta N^2} \mkern-4mu \sum_{ i,j} \mkern-3mu t_{ij} e^{i\mathbf{q}\cdot \mathbf{r}_{ij}} \mkern-4mu \left| \mkern-2mu \sum_\mathbf{k} \mkern-4mu e^{i\mathbf{k}\cdot\mathbf{r}_{ij}} \mkern-2mu \langle n_{\mathbf{k}\uparrow}\rangle  \mkern-3mu \right|^2, 
    \label{eq:ska_solution}
    \vspace{-0.1in}
\end{equation}
where $\mu = -\frac{E_o}{N\delta}$ and $\epsilon_\mathbf{q}=-\frac{1}{N}\sum_{ i,j} t_{ij} e^{i\mathbf{q}\cdot\mathbf{r}_{ij}}$  with $\mathbf{r}_{ij}=\mathbf{r}_i-\mathbf{r}_j$. 
For isotropic case, $t_{ij}=t$, the last integral term simplifies to $-\epsilon_\mathbf{q}\delta \left(\frac{\mu}{zt}\right)^2$, where $z$ is the coordination number. 
Since the six-fold rotational symmetry is broken in our anisotropic case, we solve Eq.~\eqref{eq:ska_solution} numerically. 
The first two terms, $\mu-\epsilon_{\mathbf{k}_F}\geq 0$, capture the net energy cost for up electrons needed to avoid the inserted down electron.
The remaining last two terms, proportional to $\epsilon_\mathbf{q}$, account for the energy gain from the delocalized down electron.

To find the minimum excitation energy, we replace $\epsilon_\mathbf{q}$ by the band bottom energy $\epsilon_\mathbf{q=0}$, as band minima occurs at $\mathbf{q=0}$ for all $t'>0$.
The interplay between hole density $\delta$ and $t'$ modifies the magnon's effective band curvature. 
When either parameter increases, the band steepens, eventually leading to the ``$k_F$ instability" of NFM, as illustrated in Fig.~\ref{fig:multi_hole}(b).
At $\delta=0.05$, the instability occurs around $t'\approx0.57t$, which is higher than our estimated critical value $t'_c\approx0.42t$ for a single hole in the thermodynamic limit. 
This discrepancy represents the inadequacy of the mean-field SKA ansatz in incorporating the exact correlation effects among constituents. 
Further increasing $\delta$ enhances the curvature, shifting the instability to lower $t'$, and NFM ultimately vanishes below $\delta\approx0.49$.

\textit{Conclusions}--We have explored the instability of Nagaoka ferromagnet and the resulting quantum phase transition by controlling dopants' kinetic frustration via tuning the lattice geometry between square and triangular limits. Combining analytical insights with large-scale numerical simulations, we have identified the microscopic origin of this instability both in the single-hole limit and at finite hole density.
Since Nagaoka ferromagnetism at finite doping is stabilized by a macroscopic kinetic energy gain~\cite{becca2001nagaoka}, it remains robust against moderate perturbations such as kinetic frustration ($t'/t$) and strong on-site repulsion ($U$), establishing it as a realistic many-body phenomenon rather than just a pathological feature of the single-hole limit.
Beyond the Nagaoka regime, signatures of ground-state quantum phase transitions--identified through local magnetic correlations--indicate that competing magnetic tendencies yield critical behavior with a spiral spin density phase that connects the Nagaoka and Haerter-Shastry limits. These local magnetic correlations and the critical role of kinetic frustration could be directly probed in future cold-atom experiments and moiré Hubbard platforms. In such settings, our analysis can be extended to a finite but large-$U$ regime on tunable square-triangular lattices, where itinerant holes locally suppress the superexchange mechanism. 

\textit{Acknowledgments}--- We thank S. Sherif for discussions on related work. This work is primarily supported by the U.S. Department of Energy (DOE), Office of Science, Basic Energy Sciences, under Early Career Award No.~DE-SC0024524. 
 HJC acknowledges funding from the National Science Foundation (NSF) Grant No. DMR 2046570 under the CAREER program, and the National High Magnetic Field Laboratory (NHMFL). The NHMFL is supported by NSF/DMR-2128556 and the state of Florida. 
 YP is supported by NSF Grants  No.\ PHY-2216774 and No.\ DMR-2406524. DNS is supported by the U.S. DOE, Office of Basic Energy Sciences under Grant No. DE-FG02-06ER46305 for large-scale simulations of interacting systems.  We used the Pascal computing resource at EU, yquantum at CSUN, and Planck at FSU for numerical simulations.
 
\bibliography{references_formatted}

\twocolumngrid 

\clearpage
\onecolumngrid

\begin{center}
  \textbf{\large Supplementary Material for ``Nagaoka Instability and Quantum Phase Transition via Kinetic Frustration Control''}

\vspace{10pt}

Prakash Sharma$^{1,2,*}$, Yang Peng$^{2,3}$, Donna N. Sheng$^2$, Hitesh J. Changlani$^{3,4}$ and Yao Wang$^1$ \\
\textit{$^1$Department of Chemistry, Emory University, Atlanta, Georgia 30322, USA\\
$^2$Department of Physics and Astronomy, California State University, Northridge, California 91330, USA\\
$^3$Institute of Quantum Information and Matter and Department of Physics,\\ California Institute of Technology, Pasadena, CA 91125, USA\\
$^4$Department of Physics, Florida State University, Tallahassee, Florida 32306, USA\\
$^5$National High Magnetic Field Laboratory, Tallahassee, Florida 32310, USA}
\end{center}
\vspace{20pt}

\renewcommand{\thesection}{S\arabic{section}}
\setcounter{section}{0}
\renewcommand{\thefigure}{S\arabic{figure}}
\setcounter{figure}{0}
\renewcommand{\theequation}{S\arabic{equation}}
\setcounter{equation}{0}
\renewcommand{\thetable}{S\arabic{table}}

\allowdisplaybreaks

\vspace{-0.15in}
\section{Ferromagnetism in a triangular geometry}
\begin{figure*}
    \includegraphics[]{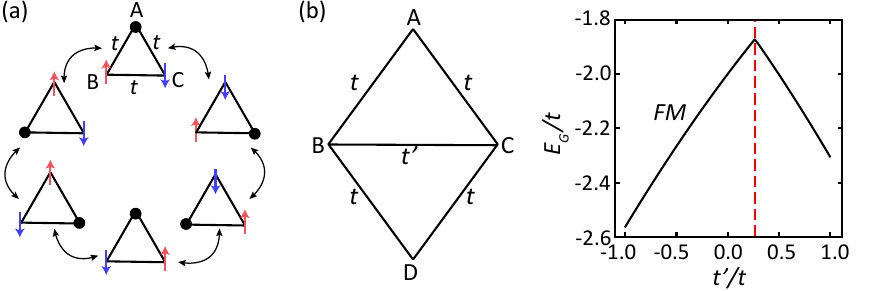}
     \caption{\label{fig:connected graph} (a) The six possible configurations of one hole (represented as a solid dot) and two spins on a triangle forming a fully connected graph: any two successive configurations are connected by a non-vanishing matrix element. Vertices A, B, and C  denote the three sublattices. (b) (Left) A rotated square geometry with a diagonal hopping $t'$. (Right) Ground state energy on this four-site geometry with one hole and three spins as a function of kinetic frustration ratio $t'/t$ .}
     \label{fig:SM_fig_1}
\end{figure*}

\vspace{-0.12in}
Nagaoka ferromagnetism was originally demonstrated for a single hole dopant at $U=\infty$ on bipartite lattices~[1]~, which Tasaki later generalized to broader graph structures~[2]. Beyond the conditions of a single hole and infinitely strong on-site repulsion, Tasaki's generalization requires two additional necessary conditions for the unique existence of the ferromagnetic ground state: (i) each hopping matrix element must be non-positive (under the sign convention used for the Hamiltonian in the main text), and (ii) the configuration space must be fully connected. To apply this theorem, let us first consider a simple case: a triangular plaquette containing one hole and two electrons, which forms a fully connected graph, as illustrated in Fig.~\ref{fig:SM_fig_1}(a). We then choose a basis set $\{\ket{\phi_i}\}$ with a phase convention in which the orbital phases are negative when the hole is on sublattices A or C, and positive when it is on sublattice B (this is merely a gauge choice and can be made in any order):
\begin{equation}
    \ket{\phi_i} \equiv \{-\ket{0,\sigma,\sigma'}, \ket{\sigma,0,\sigma'}, -\ket{\sigma,\sigma',0} \},
\end{equation}
where $\sigma$ and $\sigma'=\{\uparrow,\downarrow\}$.
With this gauge choice, the $U=\infty$ Hubbard model
\begin{equation}
    \mathcal{H} = -\sum_{\langle i,j \rangle,\sigma} t_{ij} c_{i\sigma}^\dagger c_{j\sigma} \left(1-n_{j-\sigma}\right)
    \left(1-n_{i-\sigma} \right) + \text{h.c.},
    \label{eq:Ham}
    \vspace{-0.08in}
\end{equation}
with $t_{ij}=t>0$ takes non-negative form. Restricted to total $S_z=0$ sector, we get
\begin{equation}
H =
\left[
\begin{array}{
  c @{\hskip 10pt} c @{\hskip 10pt} c @{\hskip 10pt}
  c @{\hskip 10pt} c @{\hskip 10pt} c
}
0 & 0 & t & 0 & 0 & t \\
0 & 0 & 0 & t & t & 0 \\
t & 0 & 0 & 0 & t & 0 \\
0 & t & 0 & 0 & 0 & t \\
0 & t & t & 0 & 0 & 0 \\
t & 0 & 0 & t & 0 & 0
\end{array}
\right].
\label{eq:Ham_matrix}
\end{equation}
This form violates Tasaki's condition (i) above.
However, since $H$ is irreducible and non-negative, the Perron-Frobenius theorem implies that there is a unique eigenstate corresponding to the largest eigenvalue $E_{max}$, which is a uniform superposition of all $\ket{\phi_i}$ with strictly positive coefficients. This is a ferromagnetic state 
\[ \ket{\psi_t}=
\frac{1}{\sqrt{3}} \bigg(
\begin{tikzpicture}[scale=1.5, baseline=(current bounding box.center)]
    \coordinate (A) at (0, 0.4);
    \coordinate (B) at (-0.3, 0);
    \coordinate (C) at (0.3, 0);
    
    \draw[thick] (A) -- (B) -- (C) -- cycle;

    \fill[black] (A) circle (2pt);

    \begin{scope}
        \fill[gray!30] (0,0) ellipse [x radius=0.4, y radius=0.1];
    \end{scope}

    \draw[thick, ->] (B) ++(0, -0.09) -- ++(0, 0.25); 
    \draw[thick, ->] (C) ++(0, -0.09) -- ++(0, 0.25); 
\end{tikzpicture}
+
\begin{tikzpicture}[scale=1.5, baseline=(current bounding box.center)]
    \coordinate (A) at (0, 0.4);
    \coordinate (B) at (-0.3, 0);
    \coordinate (C) at (0.3, 0);
    
    \draw[thick] (A) -- (B) -- (C) -- cycle;

    \fill[black] (B) circle (2pt);

    \begin{scope}
        \fill[gray!30,rotate around={-49:(0.32,-0.05)}] (0, -0.05) ellipse [x radius=0.35, y radius=0.1];
    \end{scope}

    \draw[thick, ->] (A) ++(0, -0.15) -- ++(0, 0.25); 
    \draw[thick, ->] (C) ++(-0.01, -0.09) -- ++(0, 0.25); 
\end{tikzpicture}
+
\begin{tikzpicture}[scale=1.5, baseline=(current bounding box.center)]
    \coordinate (A) at (0, 0.4);
    \coordinate (B) at (-0.3, 0);
    \coordinate (C) at (0.3, 0);
    
    \draw[thick] (A) -- (B) -- (C) -- cycle;

    \fill[black] (C) circle (2pt);

    \begin{scope}
        \fill[gray!30,rotate around={49:(-0.3,-0.05)}] (0, 0) ellipse [x radius=0.35, y radius=0.1];
    \end{scope}

    \draw[thick, ->] (A) ++(0, -0.09) -- ++(0, 0.25); 
    \draw[thick, ->] (B) ++(0, -0.09) -- ++(0, 0.25); 
\end{tikzpicture}
\bigg)
\]
with eigenvalue $E_{max}=2t$. Inside the gray region, the two parallel spins represent a triplet state $\frac{1}{\sqrt{2}}(\ket{\uparrow \downarrow}+ \ket{\downarrow \uparrow})$. However, this ferromagnetic state is not the ground state but the highest excited state.
The ground state of the Hamiltonian~\eqref{eq:Ham_matrix} is a resonating singlet   
\[ \ket{\psi_s}=
\frac{1}{\sqrt{3}} \bigg(
\begin{tikzpicture}[scale=1.5, baseline=(current bounding box.center)]
    \coordinate (A) at (0, 0.4);
    \coordinate (B) at (-0.3, 0);
    \coordinate (C) at (0.3, 0);
    
    \draw[thick] (A) -- (B) -- (C) -- cycle;

    \fill[black] (A) circle (2pt);

    \begin{scope}
        \fill[gray!30] (0,0) ellipse [x radius=0.4, y radius=0.1];
    \end{scope}

    \draw[thick, <-] (B) ++(0, -0.09) -- ++(0, 0.25); 
    \draw[thick, ->] (C) ++(0, -0.09) -- ++(0, 0.25); 
\end{tikzpicture}
+
\begin{tikzpicture}[scale=1.5, baseline=(current bounding box.center)]
    \coordinate (A) at (0, 0.4);
    \coordinate (B) at (-0.3, 0);
    \coordinate (C) at (0.3, 0);
    
    \draw[thick] (A) -- (B) -- (C) -- cycle;

    \fill[black] (B) circle (2pt);

    \begin{scope}
        \fill[gray!30,rotate around={-49:(0.32,-0.05)}] (0, -0.05) ellipse [x radius=0.35, y radius=0.1];
    \end{scope}

    \draw[thick, <-] (A) ++(0, -0.15) -- ++(0, 0.25); 
    \draw[thick, ->] (C) ++(-0.01, -0.09) -- ++(0, 0.25); 
\end{tikzpicture}
+
\begin{tikzpicture}[scale=1.5, baseline=(current bounding box.center)]
    \coordinate (A) at (0, 0.4);
    \coordinate (B) at (-0.3, 0);
    \coordinate (C) at (0.3, 0);
    
    \draw[thick] (A) -- (B) -- (C) -- cycle;

    \fill[black] (C) circle (2pt);

    \begin{scope}
        \fill[gray!30,rotate around={49:(-0.3,-0.05)}] (0, 0) ellipse [x radius=0.35, y radius=0.1];
    \end{scope}

    \draw[thick, <-] (A) ++(0, -0.09) -- ++(0, 0.25); 
    \draw[thick, ->] (B) ++(0, -0.09) -- ++(0, 0.25); 
\end{tikzpicture}
\bigg)
\]
with energy $E_s = -2t$. This also connects to the fact that for $t>0$, a hole is frustrated in the ferromagnetic background, and it completely releases its kinetic frustration by hopping in a singlet background.
This scenario can be reversed by flipping the sign of the hopping term. If we choose $t<0$, the ground state and the highest excited state swap, thereby establishing FM as a unique ground state in a frustrated triangle without requiring particle-hole symmetry.
This argument holds on a 2D triangular lattice with exactly one hole and corresponds to Tasaki's generalization of Nagaoka ferromagnetism.  

We now extend this analysis to a more general case of our \( t\text{-}t' \) model by considering a four-site square (rotated by $45^0$) with diagonal hopping \( t' \) [Fig.~\ref{fig:SM_fig_1}(b) (left)]. Using a similar gauge choice, we define the basis states as
\[
\ket{\phi_i} \equiv \left\{
  -\ket{0,\sigma,\sigma',\sigma''},\ 
   \ket{\sigma,0,\sigma',\sigma''},\ 
  -\ket{\sigma,\sigma',0,\sigma''},\ 
   \ket{\sigma,\sigma',\sigma'',0}
\right\},
\]
where a negative sign is assigned when the hole is on sublattices A or D, and a positive sign when it is on B or C. With this choice, the Hamiltonian \( \mathcal{H} \) takes a non-negative form for $t,t'>0$, thereby violating condition (i) once again. The evolution of ground state energy as a function of $t'/t$ is shown in Fig.~\ref{fig:SM_fig_1}(b) (right panel), which exhibits a monotonic rise until a critical value of $t'/t\approx0.26$. Despite Tasaki's condition (i) being broken for $t>0$, the ground state remains fully spin-polarized with total spin $S=3/2$ until this critical point. This simple example illustrates that Tasaki's criterion (i) and (ii) for a unique ferromagnetic ground state are sufficient but not necessary, as we also demonstrated in the main text. For $t'<0$, a saturated ferromagnet is always expected, as we discussed in the main text. 

\section{Analytic solution for one-hole-one-magnon states}
\vspace{-0.1in}
In the main text, Eq.(2), we consider the expression for one-hole-one-magnon states
\begin{equation}
    \ket{\psi}_{1H1M} = \sum_{m \neq n} \alpha_{mn} c_{m\uparrow} S_n^- \ket{\text{FM}},
    \label{eq:1h1m}
\end{equation}
where $\ket{\text{FM}}=\prod_{i=1}^N c_{i\uparrow}^\dagger \ket{0}$ and $\alpha_{mn}$ is the amplitude of having a hole at position $m$ and a spin-flip at $n$. 
Acting Eq.~\eqref{eq:Ham} on Eq.~\eqref{eq:1h1m}, we get
\begin{equation}   \mathcal{H}\ket{\psi}_{1H1M}=\sum_{\langle i,j\rangle}t_{ij} \left[\alpha_{ji}\ket{ij} + \alpha_{ij}\ket{ji}\right] + \sum_{\langle i,j\rangle}t_{ij}\sum_n \left[\alpha_{jn}\ket{in}+\alpha_{in}\ket{jn} \right],
\label{eq:Ham_expanded}
\end{equation}
where $\ket{ij}=c_{i\uparrow}S_j^-\ket{\text{FM}}$. Assuming $\ket{ls}$ is one of the orthogonal states in the superposition, then by projecting Eq.~\eqref{eq:Ham_expanded} onto $\ket{ls}$, i.e., 
\vspace{-0.1in}
$$\bra{ls}\mathcal{H}\ket{\psi}_{1H1M}=E\bra{ls}\psi\rangle_{1H1M},$$ we get,
\vspace{-0.1in}
\begin{equation}
    t_{sl}\alpha_{sl}\sum_{\boldsymbol{\delta}}\left[\delta(l-s-\boldsymbol{\delta})+\delta(l-s+\boldsymbol{\delta}) \right] +\sum_{\boldsymbol{\delta}}\left[ t_{l+\boldsymbol{\delta},l}\alpha_{l+\boldsymbol{\delta},s}  + t_{l-\boldsymbol{\delta},l}\alpha_{l-\boldsymbol{\delta},s}\right]=E \alpha_{ls},
    \label{eq:projected_H}
\end{equation}
where the sum over $\boldsymbol{\delta}$ runs over the set of nearest-neighbor vectors. For example, in the case of a triangular lattice, $\boldsymbol{\delta}$ takes three values:
$\boldsymbol{\delta}_1=(1,0)\mathbf{a}$, $\boldsymbol{\delta}_2=(-1/2,\sqrt{3}/2)\mathbf{a}$, and $\boldsymbol{\delta}_3 = (-1/2,-\sqrt{3}/2)\mathbf{a}$, where $\mathbf{a}$ denoting the lattice constant.

Now we separate the coordinates into center of mass and relative frame as $\mathbf{R}= \mathbf{r}_s$ and $\mathbf{r}=\mathbf{r}_l-\mathbf{r}_s$, and write the coefficients  
$\alpha_{ls}=\alpha(\mathbf{R+r},\mathbf{R})$.
Eq.~\eqref{eq:projected_H} takes following form 
\begin{equation}
    t_{\mathbf{r}}\, \alpha( \mathbf{R},\mathbf{R} + \mathbf{r}) 
    \sum_{\boldsymbol{\delta}} \left[ 
        \delta(\mathbf{r} + \boldsymbol{\delta}) + 
        \delta(\mathbf{r} - \boldsymbol{\delta}) 
    \right] + \sum_{\boldsymbol{\delta}}t_{\boldsymbol{\delta}} \left[\alpha(\mathbf{R+r}+\boldsymbol{\delta},\mathbf{R}) +  \alpha(\mathbf{R+r}-\boldsymbol{\delta}, \mathbf{R}) \right] = E\alpha(\mathbf{R+r},\mathbf{R})
    \label{eq:relative_H}
\end{equation}
Using periodic boundary conditions to find the solution corresponding to a specific center of mass momentum, we introduce the Fourier transform:
\begin{equation}
    \alpha(\mathbf{\mathbf{R+r},\mathbf{R}})=\sum_\mathbf{P} \psi_{\mathbf{P}}(\mathbf{r})e^{i\mathbf{P\cdot R}}.
\end{equation}
Eq.~\eqref{eq:relative_H} becomes
\vspace{-0.08in}
\begin{equation}
    t_{\mathbf{r}}\, \psi_{\mathbf{P}}(-\mathbf{r}) 
    \sum_{\boldsymbol{\delta}} \left[
        e^{i \mathbf{P} \cdot \boldsymbol{\delta}}\, \delta(\mathbf{r} - \boldsymbol{\delta}) +
        e^{-i \mathbf{P} \cdot \boldsymbol{\delta}}\, \delta(\mathbf{r} + \boldsymbol{\delta})
    \right] + \sum_{\boldsymbol{\delta}} t_{\boldsymbol{\delta}}\left[ \psi_{\mathbf{P}}(\mathbf{r}+\boldsymbol{\delta}) + \psi_{\mathbf{P}}(\mathbf{r}-\boldsymbol{\delta}) \right]=E\psi_\mathbf{P}(\mathbf{r})
    \label{eq:relative_H2}
\end{equation}
We can rewrite the last equation into compact form
\begin{equation}
    \sum_\mathbf{r'} h_\mathbf{P}(\mathbf{r,r'})\psi_\mathbf{P}(\mathbf{r'}) = E\psi_\mathbf{P}(\mathbf{r}),
\end{equation}
where $h_\mathbf{P}$ is an effective tight-binding Hamiltonian for the hole-magnon system, given by
\begin{equation} h_\mathbf{P}(\mathbf{r},\mathbf{r}^\prime)= \sum_\mathbf{\boldsymbol{\delta}
}\left( t_\mathbf{\boldsymbol{\delta}
} \left[ e^{i\mathbf{P.\boldsymbol{\delta}
}} \delta(\mathbf{r-\boldsymbol{\delta}
}) + e^{-i\mathbf{P.\boldsymbol{\delta}
}} \delta(\mathbf{r+\boldsymbol{\delta}
}) \right]\delta(\mathbf{r^\prime+\mathbf{r}}) 
    + t_\mathbf{\boldsymbol{\delta}
}\left[\delta(\mathbf{r^\prime-r-\boldsymbol{\delta}
})+\delta(\mathbf{r^\prime-r+\boldsymbol{\delta}
}) \right]
    \right),
\label{eq:tb_H}
\end{equation}
For convenience, we rewrite Eq.~\eqref{eq:relative_H2} in the following short form
\begin{equation}
E\psi_{\boldsymbol{P}}(\boldsymbol{r})-\sum_{\pm\boldsymbol{\delta}}t_{\boldsymbol{\delta}}\psi_{\boldsymbol{P}}(\boldsymbol{r}+\boldsymbol{\delta})=\sum_{\pm\boldsymbol{\delta}}t_{\boldsymbol{\delta}}\psi_{\boldsymbol{P}}(-\boldsymbol{r})e^{i\boldsymbol{P}\cdot\boldsymbol{\delta}}\delta(\boldsymbol{r}-\boldsymbol{\delta})
\label{eq:compact}
\end{equation}
which is solved with the condition $\psi_{P}(0)=0$. After performing another Fourier transform 
\begin{equation}
\psi_{\boldsymbol{P}}(\boldsymbol{r})=\frac{1}{V}\sum_{\boldsymbol{q}}\varphi_{\boldsymbol{P}}(\boldsymbol{q})e^{i\boldsymbol{q}\cdot\boldsymbol{r}},
\end{equation}
the RHS of Eq.~\eqref{eq:compact} can be rewritten as
\begin{equation}
\frac{1}{V}\sum_{\boldsymbol{q}}e^{i\boldsymbol{q}\cdot\boldsymbol{r}}\left(E-\sum_{\pm\boldsymbol{\delta}}t_{\boldsymbol{\delta}}e^{i\boldsymbol{q}\cdot\boldsymbol{\delta}}\right)\varphi_{\boldsymbol{P}}(\boldsymbol{q})=\sum_{\pm\boldsymbol{\delta}}t_{\boldsymbol{\delta}}\psi_{\boldsymbol{P}}(\boldsymbol{-}\boldsymbol{r})e^{i\boldsymbol{P}\cdot\boldsymbol{\delta}}\delta(\boldsymbol{r}-\boldsymbol{\delta})
\end{equation}
Multiplying $e^{-i\boldsymbol{k}\cdot\boldsymbol{r}}$ on both sides
and integrate over $\boldsymbol{r}$, we obtain
\begin{equation}
\left(E-\sum_{\pm\boldsymbol{\delta}}t_{\boldsymbol{\delta}}e^{i\boldsymbol{k}\cdot\boldsymbol{\delta}}\right)\varphi_{\boldsymbol{P}}(\boldsymbol{k})=\sum_{\pm\boldsymbol{\delta}}t_{\boldsymbol{\delta}}\psi_{\boldsymbol{P}}(-\boldsymbol{\delta})e^{i(\boldsymbol{P-k})\cdot\boldsymbol{\delta}}.
\end{equation}
Thus, 
\begin{equation}
\varphi_{\boldsymbol{P}}(\boldsymbol{k})=\left(E-\sum_{\pm\boldsymbol{\delta}}t_{\boldsymbol{\delta}}e^{i\boldsymbol{k}\cdot\boldsymbol{\delta}}\right)^{-1}\sum_{\pm\boldsymbol{\delta}}t_{\boldsymbol{\delta}}\psi_{\boldsymbol{P}}(-\boldsymbol{\delta})e^{i(\boldsymbol{P-k})\cdot\boldsymbol{\delta}}.
\end{equation}
We thus have
\begin{equation}
\psi_{\boldsymbol{P}}(\boldsymbol{r})=\frac{1}{V}\sum_{\boldsymbol{k}}\frac{e^{i\boldsymbol{k}\cdot\boldsymbol{r}}}{E-\sum_{\pm\boldsymbol{\delta}}t_{\boldsymbol{\delta}}e^{i\boldsymbol{k}\cdot\boldsymbol{\delta}}}\sum_{\pm\boldsymbol{\delta}}t_{\boldsymbol{\delta}}\psi_{\boldsymbol{P}}(-\boldsymbol{\delta})e^{i(\boldsymbol{P-k})\cdot\boldsymbol{\delta}}.
\end{equation}
At $\boldsymbol{P}=0$, we have
\begin{equation}
\psi(\boldsymbol{r})=\frac{1}{V}\sum_{\boldsymbol{k}}\frac{e^{i\boldsymbol{k}\cdot\boldsymbol{r}}}{E-\sum_{\pm\boldsymbol{\delta}}t_{\boldsymbol{\delta}}e^{i\boldsymbol{k}\cdot\boldsymbol{\delta}}}\sum_{\pm\boldsymbol{\delta}}t_{\boldsymbol{\delta}}e^{-i\boldsymbol{k}\cdot\boldsymbol{\delta}}\psi(-\boldsymbol{\delta}).
\label{eq:solution}
\end{equation}
where we introduced the notation $\psi_{\boldsymbol{P}=0}\equiv\psi$.
Further, we will consider odd solutions, namely $\psi(\mathbf{r})=-\psi(-\mathbf{r})$, as only odd solutions yield bound states, as discussed in Ref.~[3]~\cite{davydova2023itinerant}. 
Also, notice that Eq.~\eqref{eq:1h1m} through Eq~\eqref{eq:solution} are valid in any spatial dimension. 
\begin{figure}
    \centering
    \includegraphics[width=\linewidth]{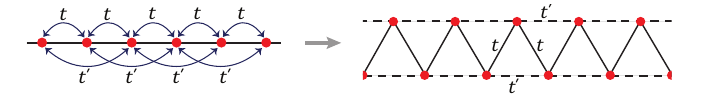}
    \caption{Frustrated 1D chain with nearest-neighbor hopping amplitude \( t \) and next-nearest-neighbor hopping \( t' \). This geometry maps onto a zigzag ladder, with hopping along horizontal bonds given by \( t' \) and zigzag bonds by \( t \), making it a minimal model for studying geometric frustration effects.}

    \label{fig:1D_chain}
\end{figure}
\subsection{Frustrated 1D chain}
We first consider a 1D frustrated chain with nearest neighbor hopping $t$ and next-nearest neighbor hopping $t^\prime$.
This configuration is equivalent to a zigzag ladder with hopping strength $t^\prime$ along horizontal bonds and $t$ otherwise, as shown in Fig.~\ref{fig:1D_chain}.
Strictly speaking, there is no Nagaoka ferromagnet in a 1D chain as the Hilbert space for a single hole at $t'=0$ is fragmented \,\cite{nagaoka1966ferromagnetism,tasaki1989extension}, thereby violating the connectivity condition required for the uniqueness of the FM ground state in Tasaki's theorem. Nevertheless, Tasaki's Theorem 1 in Ref.~[2] ensures the FM state still remains an exact (though degenerate) ground state.
To understand its stability against kinetic frustration $t'$, we solve Eq.~\eqref{eq:solution} in 1D limit.

In 1D chain, $\boldsymbol{\delta}\in\{a,2a\}$.
Denote $\psi_{1}\equiv\psi(a)$, $\psi_{2}\equiv\psi(2a)$, $t\equiv t_{\pm a}$,
and $t'\equiv t_{\pm2a}'$, we obtain
\begin{equation}
\psi_{1}=\frac{2}{V}\sum_{\boldsymbol{k}}\frac{ie^{ika}(t\sin ka\psi_{1}+t'\sin2ka\psi_{2})}{E-\epsilon_{k}}=-\frac{2}{V}\sum_{\boldsymbol{k}}\frac{(t\sin^{2}ka\psi_{1}+t'\sin ka\sin2ka\psi_{2})}{E-\epsilon_{k}}
\end{equation}
\begin{equation}
\psi_{2}=\frac{2}{V}\sum_{\boldsymbol{k}}\frac{ie^{i2ka}(t\sin ka\psi_{1}+t'\sin2ka\psi_{2})}{E-\epsilon_{k}}=-\frac{2}{V}\sum_{\boldsymbol{k}}\frac{(t\sin ka\sin2ka\psi_{1}+t'\sin^{2}2ka\psi_{2})}{E-\epsilon_{k}}
\end{equation}
where $\epsilon_{k}=2(t\cos ka+t'\cos2ka)$. In the above equations, the second equality is due to the fact that odd functions in $\boldsymbol{k}$ do not contribute to the summation. Thus, we obtain the condition
for bound states (set the lattice constant $a=1$, and the number
of sites is $V\equiv N$)
\begin{equation}
\det\left(\begin{array}{cc}
1+\frac{1}{N}\sum_{k}\frac{2t\sin^{2}k}{E-\epsilon_{k}} & \frac{1}{N}\sum_{k}\frac{2t'\sin k\sin2k}{E-\epsilon_{k}}\\
\frac{1}{N}\sum_{k}\frac{2t\sin k\sin2k}{E-\epsilon_{k}} & 1+\frac{1}{N}\sum_{k}\frac{2t'\sin^{2}2k}{E-\epsilon_{k}}
\end{array}\right)=0.
\end{equation}
We obtain 
\begin{equation}
\left(1+\frac{1}{N}\sum_{k}\frac{2t\sin^{2}k}{E-\epsilon_{k}}\right)\left(1+\frac{1}{N}\sum_{k}\frac{2t'\sin^{2}2k}{E-\epsilon_{k}}\right)=4tt'\left(\frac{1}{N}\sum_{k}\frac{\sin k\sin2k}{E-\epsilon_{k}}\right)^{2}.
\end{equation}
This equation contains bound solutions with energy $E$ below the bare hole minima for any $t^\prime > 0$. 
Therefore, unlike in 2D system, introducing an arbitrarily small $t^\prime > 0$ in 1D chain immediately lifts this degeneracy, and the FM state is destabilized in favor of a spin-polaron. This illustrates the dimensional dependence of the spin-polaron transition, with critical $t_c'=0$ in the zigzag ladder, as we discussed in the main text.

\begin{figure*}[ht!]
    \includegraphics[]{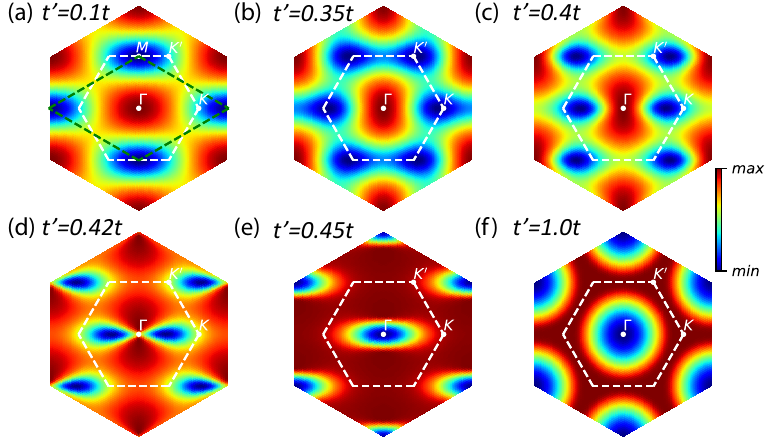}
    \caption{ The lowest energy distribution $E_{hs}(\mathbf{P})$ of the center of mass of a hole and a spin-flip obtained on $50\times 50$ sites torus at diagonal hopping strengths (a) $t'=0.1$, (b) $t'=0.35$, (c) $t'=0.4$, (d)$t'=0.42$, (e) $t'=0.45$, and (f) $t'=1.0$. With increasing $t'$, the minima of dispersion gradually shift from the $\mathbf{M}$ point towards the $\mathbf{\Gamma}$ point. $t'\approx0.42$ is the critical point where the Nagaoka ferromagnet becomes unstable and spin-polaron bound state formation starts to happen.}
    \label{fig:1h1m_dispersion}
\end{figure*}

\subsection{2D anisotropic triangular lattice}

Now we consider a 2D anisotropic triangular lattice, with $\boldsymbol{\delta}\in\{\boldsymbol{\delta_{1}},\boldsymbol{\delta}_{2},\boldsymbol{\delta}_{3}\}$.
We denote $\psi_{i}=\psi(\boldsymbol{\delta}_{i})$, $t=t_{\boldsymbol{\delta_{1}}}=t_{\boldsymbol{\delta_{2}}}$,
$t'=t_{\boldsymbol{\delta}_{3}}$. Note that $\boldsymbol{\delta}_{1,2,3}$
are not all independent, they are constrained to $\sum_{i}\boldsymbol{\delta}_{i}=0$.
Similar to the 1D case, we have
\begin{equation}
\psi_{1}=\frac{2}{V}\sum_{\boldsymbol{k}}\frac{ie^{ik_{1}}(t\sin k_{1}\psi_{1}+t\sin k_{2}\psi_{2}+t'\sin k_{3}\psi_{3})}{E-\epsilon_{\boldsymbol{k}}}=-\frac{2}{V}\sum_{\boldsymbol{k}}\frac{(t\sin^{2}k_{1}\psi_{1}+t\sin k_{1}\sin k_{2}\psi_{2}+t'\sin k_{1}\sin k_{3}\psi_{3})}{E-\epsilon_{\boldsymbol{k}}}
\end{equation}
\begin{equation}
\psi_{2}=\frac{2}{V}\sum_{\boldsymbol{k}}\frac{ie^{ik_{2}}(t\sin k_{1}\psi_{1}+t\sin k_{2}\psi_{2}+t'\sin k_{3}\psi_{3})}{E-\epsilon_{\boldsymbol{k}}}=-\frac{2}{V}\sum_{\boldsymbol{k}}\frac{(t\sin k_{1}\sin k_{2}\psi_{1}+t\sin^{2}k_{1}\psi_{2}+t'\sin k_{1}\sin k_{3}\psi_{3})}{E-\epsilon_{\boldsymbol{k}}}
\end{equation}
\begin{equation}
\psi_{3}=\frac{2}{V}\sum_{\boldsymbol{k}}\frac{ie^{ik_{3}}(t\sin k_{1}\psi_{1}+t\sin k_{2}\psi_{2}+t'\sin k_{3}\psi_{3})}{E-\epsilon_{\boldsymbol{k}}}=-\frac{2}{V}\sum_{\boldsymbol{k}}\frac{(t\sin k_{3}\sin k_{1}\psi_{1}+t\sin k_{3}\sin k_{1}\psi_{2}+t'\sin^{2}k_{3}\psi_{3})}{E-\epsilon_{\boldsymbol{k}}}
\end{equation}
where
\vspace{-0.1in}
\begin{equation}
\epsilon_{\boldsymbol{k}}=2t\left[\cos k_{1}+k_{2}\right]+2t'\cos(k_{1}+k_{2}),\quad k_{i}=\boldsymbol{k}\cdot\boldsymbol{\delta_{i}}.
\end{equation}
We also have $\delta_{3}=-\delta_{1}-\delta_{2}$. The system has
a 2-fold mirror symmetry, along the reflection direction along $\boldsymbol{\delta}_{3}$,
namely $\epsilon_{\boldsymbol{k}}$ is invariant under exchange of
$k_{1}$ and $k_{2}$. From the above equations, we have
\begin{equation}
\left(1+\frac{2t}{V}\sum_{\boldsymbol{k}}\frac{\sin k_{1}^{2}+\sin k_{1}\sin k_{2}}{E-\epsilon_{\boldsymbol{k}}}\right)(\psi_{1}+\psi_{2})+\frac{4t'}{V}\sum_{\boldsymbol{k}}\frac{\sin k_{1}\sin k_{3}}{E-\epsilon_{\boldsymbol{k}}}\psi_{3}=0
\end{equation}
\vspace{-0.1in}
\begin{equation}
\left(1+\frac{2t'}{V}\sum_{\boldsymbol{k}}\frac{\sin^{2}k_{3}}{E-\epsilon_{\boldsymbol{k}}}\right)\psi_{3}+\frac{2t}{V}\sum_{\boldsymbol{k}}\frac{\sin k_{1}\sin k_{3}}{E-\epsilon_{\boldsymbol{k}}}(\psi_{1}+\psi_{2})=0.
\end{equation}
 Thus, we obtain the condition for bound states (set the lattice constant
$a=1$, and the number of sites $N=V$)
\begin{equation}
\det\left(\begin{array}{cc}
1+\frac{2t}{N}\sum_{\boldsymbol{k}}\frac{\sin k_{1}^{2}+\sin k_{1}\sin k_{2}}{E-\epsilon_{\boldsymbol{k}}} & \frac{4t'}{N}\sum_{\boldsymbol{k}}\frac{\sin k_{1}\sin k_{3}}{E-\epsilon_{\boldsymbol{k}}}\\
\frac{2t}{N}\sum_{\boldsymbol{k}}\frac{\sin k_{1}\sin k_{3}}{E-\epsilon_{\boldsymbol{k}}} & 1+\frac{2t'}{N}\sum_{\boldsymbol{k}}\frac{\sin^{2}k_{3}}{E-\epsilon_{\boldsymbol{k}}}
\end{array}\right)=0.
\end{equation}
We obtain 
\vspace{-0.1in}
\begin{equation}
\left(1+\frac{2t}{N}\sum_{\boldsymbol{k}}\frac{\sin k_{1}^{2}+\sin k_{1}\sin k_{2}}{E-\epsilon_{\boldsymbol{k}}}\right)\left(1+\frac{2t'}{N}\sum_{\boldsymbol{k}}\frac{\sin^{2}k_{3}}{E-\epsilon_{\boldsymbol{k}}}\right)=\frac{8tt'}{N^{2}}\left(\sum_{k}\frac{\sin k_{1}\sin k_{3}}{E-\epsilon_{\boldsymbol{k}}}\right)^{2}.
\end{equation}
This is the bound state equation used in Eq.(3) of the main text. This yields bound state solution below the bare hole minima for $t'\gtrsim 0.42$, in the limit $N\rightarrow\infty$.

\section{Properties of one-hole-one-magnon state}
\begin{figure}[ht]
    \centering
    \includegraphics[width=\linewidth]{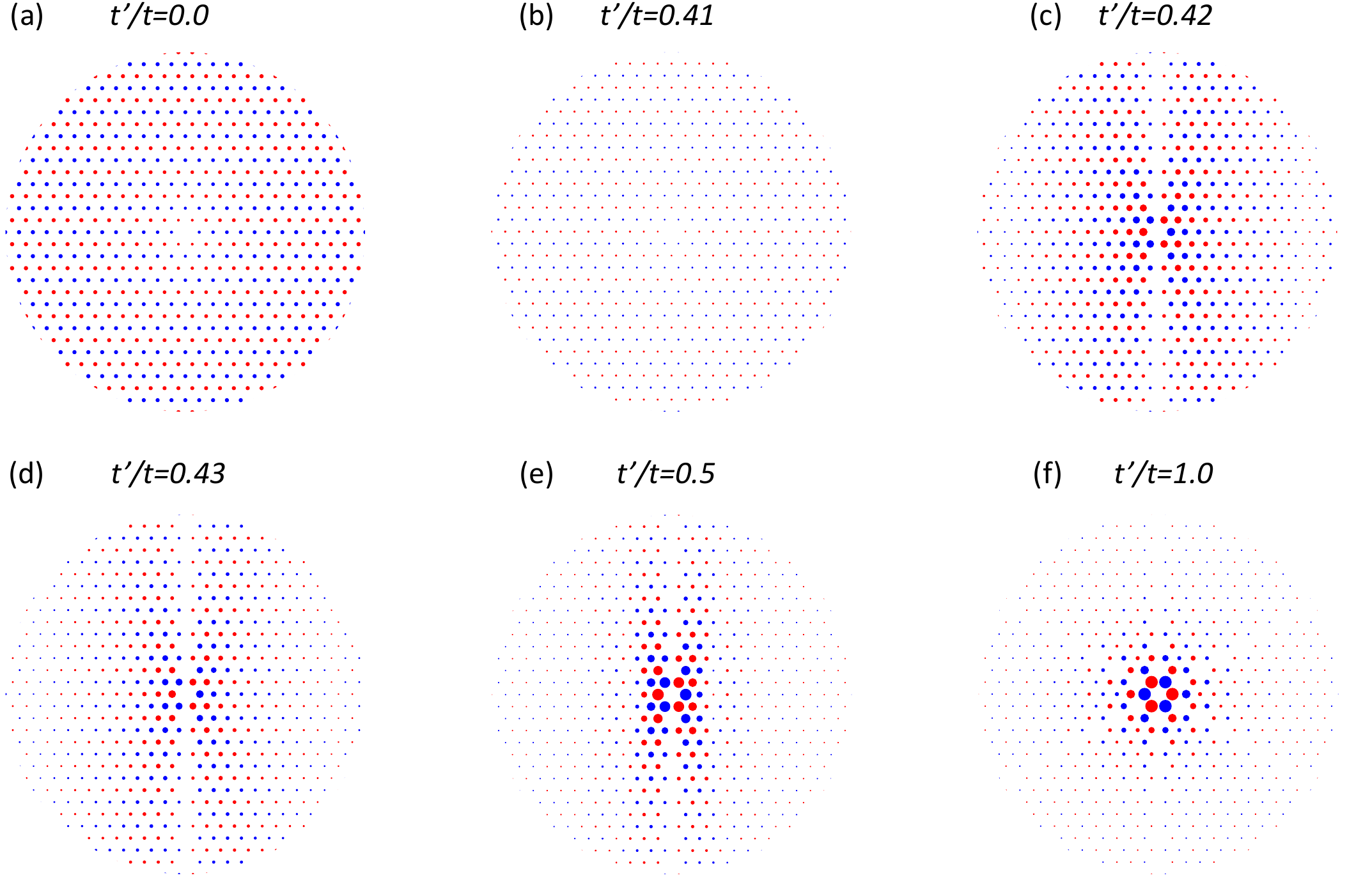}
   \caption{Real-space distribution of the wavefunction amplitudes for the one-hole–one-magnon state at $\mathbf{P=\Gamma}$. The size of the dot indicates the amplitude, while color indicates phase (red for negative, blue for positive). All panels share the same color scale to enable direct comparison. The magnitude of dots physically represents the probability of finding the magnon around the hole (which is fixed at the center of the cluster).}

    \label{fig:probability_amp}
\end{figure}

In Fig. 2(a) of the main text, we show only the low-lying eigen-energies corresponding to three specific center-of-mass momentum sectors, which separate the Nagaoka state from the spin-polaron. Here, we numerically solve the Hamiltonian Eq.~\eqref{eq:tb_H} on $50\times50$ triangular torus at various values of $t'$, and present the lowest energy distribution of a one-hole-one-magnon state in momentum space (see Fig.~\ref{fig:1h1m_dispersion}). At small frustration, $t'=0.1t$, the minima is located at $\mathbf{M}$ point, essentially forming a square lattice unit cell in the reciprocal space. As $t'$ increases, the minima gradually evolve towards the center of the Brillouin zone. At $t'=0.42t$, two minima approaching from opposite directions form a dumbbell shape touching the center at $\mathbf{\Gamma}$. At this critical point, we observe that the previously unbound hole-magnon pair forms a bound state--the spin-polaron (see Fig.~\ref{fig:probability_amp}). 
Above this critical point, the two minima merge, and the energy minimum of the spin-polaron becomes centered at $\Gamma=(0,0)$, as we mentioned in the main text.  

In the main text, we discussed that the spatial extent of the bound state decreases with increasing \( t' \), reaching its most tightly bound form in the triangular limit. Here, we directly visualize this effect by plotting the real-space distribution of the wavefunction, shown in Fig.~\ref{fig:probability_amp} for the lowest state at center-of-mass momentum \( \mathbf{\Gamma} = (0, 0) \).
Below the critical point, for example at \(t'/t=0.0\) through \( t'/t = 0.41 \), the hole–magnon pair is unbound. At the critical value \( t'/t = 0.42 \), a bound state forms, as indicated by the enhanced amplitudes around the central site. Note that the sixfold rotational symmetry is broken due to the anisotropic hopping \( t' \), unless $t'=t$. As discussed in the main text, the extent of the bound state gradually shrinks with increasing \( t' \), ultimately becoming strongly localized — with a spatial extent of roughly one or two lattice spacing — in the triangular limit \( t'/t=1.0 \).

{\color{blue}{$^*$  sharmaprakash078@gmail.com}} \\
$[\text{ 1 }]$ Y. Nagaoka, \textit{Ferromagnetism in a narrow, almost half-filled s band}, Physical Review, \textbf{147}, 392 (1966) \\
$[\text{ 2 }]$ H. Tasaki, \textit{Extension of Nagaoka’s theorem on the large-U Hubbard model}, Physical Review B, \textbf{40}, 9192 (1989) \\
$[\text{ 3 }]$ M. Davydova, Y. Zhang, and L. Fu, \textit{Itinerant spin polaron and metallic ferromagnetism in semiconductor moir\'e superlattices}, Physical Review B, \textbf{107}, 224420, (2023)

\end{document}